# A fine-grained, versatile index of remoteness to characterize place-level rurality

Johannes H. Uhl, Stefan Leyk, Lori M. Hunter, Catherine B. Talbot, Dylan S. Connor, Jeremiah J. Nieves, Myron Gutmann

## Introduction

Many demographic processes and patterns play out differently between urban and rural areas. While the understanding of rural regions is critical for policy development, knowledge of rural socioeconomic and demographic conditions in the United States is mainly derived from county-scale data analysis. However, county-level data misses the internal heterogeneity and thus, introduces bias when using such data for place-level studies, focusing on finer-grained spatial units, such as towns and villages. Researchers commonly use the county level as analytical unit for the creation of rural-urban indices, which have, themselves, been employed in studies focused on places as related to their location along a rural-urban continuum.

Examples of rural-urban classifications in the U.S. (see Fig. 1) include the commonly used ***rural-urban continuum codes*** (RUCCs) of the U.S. Department of Agriculture (USDA) (McGranahan et al 1986, Butler 1990), which identify nine classes, i.e., three metro and six nonmetropolitan county designations. Metropolitan counties are further disaggregated based on population size of a county's metro area, while nonmetropolitan counties are further classified by their degree of urbanization and adjacency to a metro area. A related classification is the ***urban-rural classification scheme for counties*** from NCHS (National Center for Health Statistics) (Ingram & Franco 2014) which is based on metropolitan and non-metropolitan county classification in combination with population thresholds, identifying six county designations. Another classification scheme is the ***index of relative rurality*** (IRR, Waldorf 2006 and Waldorf & Kim 2018), which is a continuous county-level index based on population size, density, road network distance, and built-up areas. Further measures of the rural-urban continuum include the ***USDA urban influence codes*** (Ghelfi & Parker 1997), which are based on a modified version of the USDA RUCC classification strategy (i.e., taking into account the population of the largest city within a county, rather than an aggregated urban population) yielding 9 different classes, and the ERS (Economic Research Service) ***rural-urban commuting area codes*** (Economic Research Service 2013) which make use of the U.S. census urbanized areas and urban core designations (U.S. Census Bureau 2020), in combination with census-tract level commuting flow estimates, grouping census tracts into 10 classes of commuting levels.

Such coarse-scale data potentially ignore intra-unit variations of rurality although understanding this variation is important for understanding demographic changes at scales relevant for decision-making and development planning (e.g., census-designated places and incorporated places, herein referred to as "places"). Moreover, due to changes in both county boundaries and the definition of metro and non-metro counties over time, classifications such as the RUCCs suffer from temporal inconsistencies, also due to changes in methodology. In addition to that, more complex indices such as the IRR rely on data sources that are rarely available for early points in time, such as road network data, built-up areas, or unit-level population density.

Hence, efforts to analyze demographic processes across the rural-urban continuum at the place level and over time have been impeded by the lack of spatially fine-grained and temporally consistent indicators of rural and urban places. Thus, we propose a method to derive spatially fine-grained measures of remoteness in the U.S. (and elsewhere). This method can be used to generate consistent classifications of rural and urban places over long periods of time. This approach follows, but simplifies, characteristics commonly used to define rural-urban classes (e.g., size, distance, and spatial relationships between populated places). Specifically, the remoteness measure is based on population size of places and the weighted (Euclidean) distances to other places of different size categories. This approach relies solely on public-domain data and allows the derivation of various distance and population-based attributes. The method also allows for systematic adjustment of criteria and distance weights at the local scale to methodologically and substantively examine implications of shifts in elements of the remoteness scale.

Such refined indicators open new avenues for the study of rural development, infrastructure access, and social and economic well-being, inquiries that have been constrained to the scale of county and/or census tract by the coarse granularity of standard measures of rurality. Focusing on the spatial distributions of the remoteness index allow critical reflection on the varying rural-urban nature of places within counties or tracts, which has important implications for planning and policy. Across time, trends of such remoteness indices provide unprecedented insights into the development history of the rural United States, which can also be used for planning and prioritizing economic growth and public health in the future. Furthermore, this effort is motivated by the within-county /-tract heterogeneity being missed by county- or tract-scale classifications of rural and urban spaces, and aims to generate an index using publicly available data, accessible and flexible for scholars and policymakers. Throughout, we also aim to call attention to the importance of using small town data to develop an understanding of small towns, an understanding essential for the development of appropriate place-based policy.

## Data

Herein, we use U.S. census place polygon shapefiles and associated population counts from 1980, 1990, 2000, and 2010, obtained from NHGIS[1] (Manson et al. 2020), containing both, incorporated places and census-designated places (U.S. Census Bureau 1994).

## Approach

Using place-level population estimates provided as discrete geospatial locations, we design a method to model the remoteness of places across the region of interest (e.g., the U.S.). The remoteness index of a place is computed based on the population of the place of interest and the distance between that place and the nearest places of varying size categories (10,000-20,000, 20,000-50,000, 50,000-100,000, 100,000-250,000 and more than 250,000 people, herein referred to as population categories $pc$).

The remoteness index **RI** for a given place $p$, in a given year $t$ can be derived as the weighted average of the inverse of the population size $s_{p,t}$, of place $p$ in year $t$, and the distance measures $D_{pc,p,t}$, to the nearest place of population category $pc$ (Equation 1). All measures are log-transformed to achieve a uniformly distributed index despite skewed distributions of population (and potentially skewed distributions of distance measures due to the presence of extremely remote places):

$$\overline{RI}_{p,t} = \frac{1}{30}\left[15 \cdot \log(s_{p,t})^{-1} + \sum_{pc=1}^{5} w_{pc} \cdot \log(D_{pc,p,t})\right] \quad (1)$$

where $w_{pc}$ are weights for the distance measures to different population categories to allow for adjusting the influence of local versus regional population centers. More specifically, in our pilot experiments, we have tested different weighting schemes such as assigning equal weights ($W_{pc} = [3\ 3\ 3\ 3\ 3]$) to places of all size categories, as well as larger weights to distances to large places and smaller weights to distances to smaller places (e.g., $W_{pc} = [1\ 2\ 3\ 4\ 5]$). Note that the factor 1/30 is used in order to use integer weights in all weight scenarios, to facilitate interpretation.

The final index is calculated by scaling the raw index measures $\overline{RI}_{p,t}$ into the range [0,1]:

$$RI_{p,t} = \frac{\overline{RI}_{p,t} - \min(\overline{RI}_t)}{\max(\overline{RI}_t) - \min(\overline{RI}_t)} \quad (2)$$

---

[1] https://www.nhgis.org/

This computation yields values close to 0 for large places near other (large and/or small) places, and values close to 1 for small places, remote from other places. By approximating each place by a discrete point location (i.e., the place polygon centroid), rather than using its areal extent, and by modelling the distances between places using Euclidean rather than road network distances (cf. Weiss et al. 2018, Nelson et al. 2019), our approach is highly versatile and generalizable to data-scarce environments and (early) time periods, as retrospective areal place extents and multi-temporal road network data are rarely available.

**Preliminary results**

Based on the described calculations, we created maps representing the level of remoteness across the contiguous U.S., shown for the year 2010 in Fig. 1a. As can be seen, this depiction allows for deriving trends in remoteness for each place in the conterminous U.S. at finer spatial granularity than the IRR (Fig. 1b), and the commonly-used NCHS urban-rural classes (Fig. 1c), USDA RUCCs (Fig. 1d), and USDA urban influence codes (Fig. 1e). Our remoteness index is also expected to yield finer-grained results than the tract-level ERS rural-urban commuting areas (Fig. 1f), especially in rural areas where census tracts may be large. It is notable that the NCHS urban-rural classification (Fig. 1c) seems to be the most conservative approach, assigning large parts of the U.S. counties to the most rural class. Interestingly, counties in the Southwest (e.g., in Arizona, Nevada), typically characterized by large counties and sparse, highly clustered population distributions, are classified quite differently by the different approaches, indicating high levels of sensitivity of the classification methods to such extreme conditions.

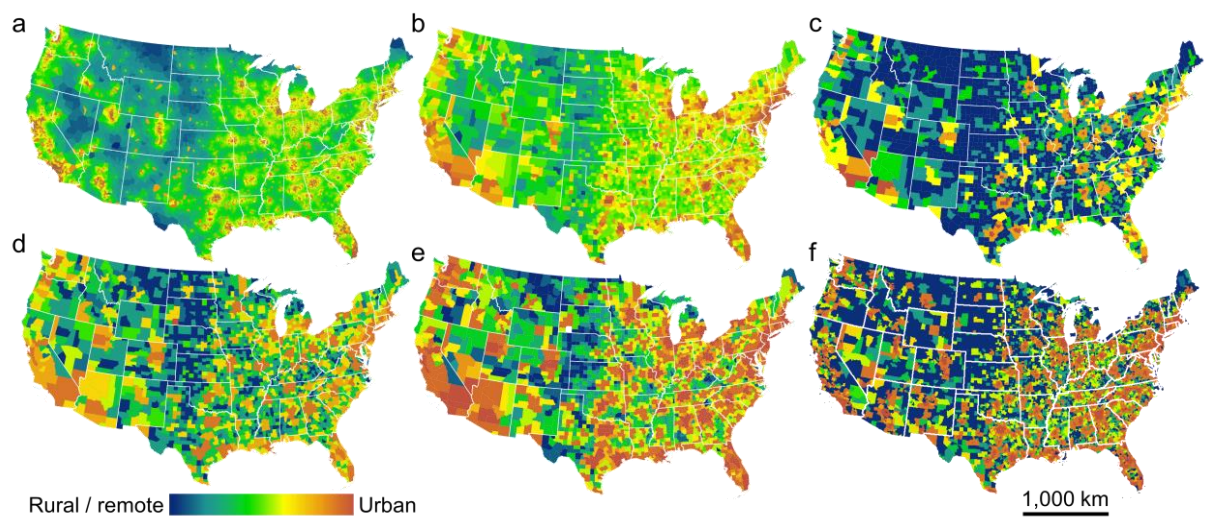

**Figure 1.** Comparison of rural-urban classification schemes: (a) Place-level remoteness in the conterminous U.S. in 2010, using an equal-weights weighting scheme. For visualization purposes, places are represented by the area of the Thiessen polygon corresponding to the place centroid. Also shown for comparison are county and tract-level rurality classifications: (b) the Index of Relative Rurality in 2010, (c) 2013 NCHS Urban–Rural Classification Scheme, (d) the USDA RUC codes in 2013, (e) .USDA urban influence codes, and (f) USDA rural-urban commuting area codes in 2010 on tract-level (ERS 2013)

These RI distributions are important ingredients for the derivation of more advanced measures of rurality (or urbanness) to better understand the rural-urban transformation in the U.S. over long time periods. Regional trends reflect differences in timing of urban development at the census place level and thus at much finer granularity than the county level data shown in Figs. 1b-f. Within-county /-tract variation of such trends will better inform decision makers as related to resource allocation. Moreover, multi-temporal versions of the remoteness index allow for a temporally consistent mapping and analysis of remoteness over time (Fig. 2).

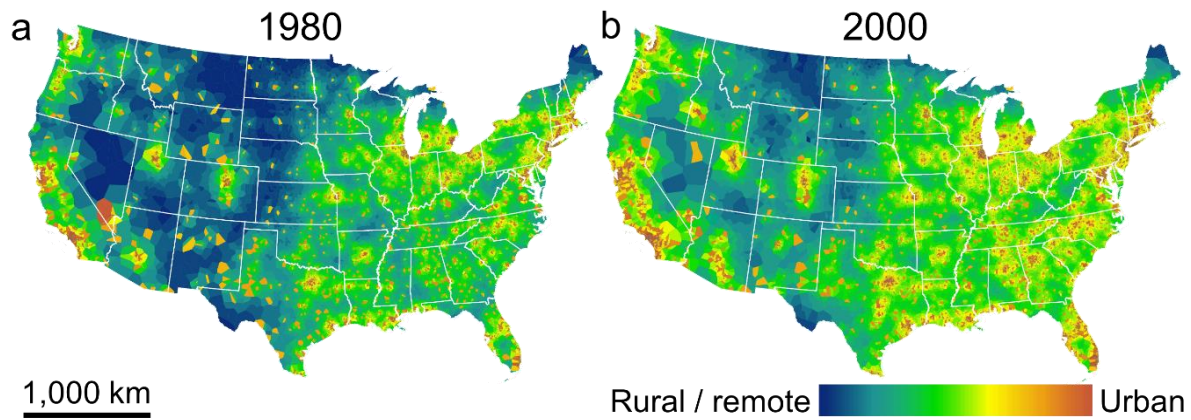

**Figure 2.** Multi-temporal depictions of the place-level remoteness index (a) in 1980 and (b) in 2000.

# Future work

We will further develop these indices, e.g. by testing the usefulness of focal (i.e., place-boundary independent) population density measures. We will also test different weighting schemes to cross-evaluate the resulting indices with other measures related to socioeconomic processes, accessibility, urban development, settlement density and intensity, as well as their relationship to existing, coarser rural-urban classifications (cf. Fig. 1). Furthermore, we will employ these measures for more refined trend analysis and shed light on regional variations in urban development to identify places of existing typical development patterns. This will be useful for planners and administrations to determine actions necessary to support communities in need and project future local and regional development.

Future steps will also include the extrapolation of the described approach to other points in time and countries in which similar data are available and accessible. Pilot experiments using harmonized U.S., Mexican and Canadian census data have shown promising results. In all, we believe this nuanced approach to rural-urban characterization using readily available decadal census data has tremendous potential for use by the demographic research community as well as policymakers.